\documentclass[aps,prb,twocolumn,superscriptaddress]{revtex4-2}

\usepackage{bm}
\usepackage{amssymb}
\usepackage{amsmath}
\usepackage{siunitx}
\usepackage{textcomp}
\usepackage{graphicx}
\usepackage{natbib}
\usepackage{color}
\usepackage{setspace}

\begin{document}

\title{Statistical laws and self-similarity of vortex rings emitted from a localized vortex tangle in superfluid ${}^4${He}}


\author{Tomo Nakagawa}
\author{Sosuke Inui}
\affiliation{Department of Physics, Osaka City University, 3-3-138 Sugimoto, 558-8585 Osaka, Japan}
\author{Makoto Tsubota}
\affiliation{Department of Physics, Osaka City University, 3-3-138 Sugimoto, 558-8585 Osaka, Japan}
\affiliation{Nambu Yoichiro Institute of Theoretical and Experimental Physics(NITEP), Osaka City University, 3-3-138 Sugimoto, 558-8585 Osaka, Japan}
\affiliation{The Advanced Research Institute for Natural Science and Technology(OCARINA),
Osaka City University, 3-3-138 Sugimoto, 558-8585 Osaka, Japan}
\author{Hideo Yano}
\affiliation{Department of Physics, Osaka City University, 3-3-138 Sugimoto, 558-8585 Osaka, Japan}
\affiliation{Nambu Yoichiro Institute of Theoretical and Experimental Physics(NITEP), Osaka City University, 3-3-138 Sugimoto, 558-8585 Osaka, Japan}


\date{\today}

\begin{abstract}
We numerically simulated quantum turbulence in superfluid $^4$He to investigate the emission of vortex rings from a localized vortex tangle. 
Turbulence is characterized by some universal statistical laws. Although there are a lot of studies on statistical laws in bulk quantum turbulence, studies in inhomogeneous or localized turbulence is scarce. We first investigate the statistical laws of localized quantum turbulence, referring to two statistical laws deduced from the vibrating wire experiments in [Yano $et$ $al.$, J. Low Temp. Phys. $\bm{196},\ 184\ (2019)$]. 
 The first law is the Poisson process for the detection of vortex rings; the vortex tangle emits vortex rings with frequencies depending on their sizes.
The second law is the power law between the frequency and the size of the emitted vortex rings, showing the self-similarity of the tangle. 
To study these statistical laws numerically, 
we developed a system similar to experiments.
First, we generate a localized statistically steady vortex tangle by injecting vortex rings from two opposite sides and causing collisions. We investigated the conditions that aid the formation of the tangles and the anisotropy of the emission of vortex rings from the tangle. Second, from the data on emitted rings, we reconstruct the two statistical laws. Results from our numerical investigations are consistent with the known self-similarity of emitted vortex rings and localized tangles.
\end{abstract}


\maketitle

\section{Introduction}
Quantum turbulence refers to turbulent states in quantum condensed fluids. It is an important phenomenon in low temperature physics and fields such as fluid mechanics, and non-equilibrium physics. Superfluid ${}^{4}$He is a typical system wherein quantum turbulence is studied. A lot of researchers have investigated superfluid $^4$He for over half a century \cite{Vinen2002,Tsubota2013,Barenghi2014,Tsubota2017}. Some statistical laws are often investigated to determine the universal properties of turbulence. In classical turbulence, an important statistical law is the Kolmogorov's law that indicates that the energy spectrum follows the $-5/3$ power law of the wave number \cite{Davidson,Frisch}. This shows self-similarity in the wave number space. In real space, the self-similarity can be expected to be the Richardson cascade wherein large-sized eddies can split into smaller sizes \cite{Davidson,Frisch}. Eddies or vortices can be responsible for the self-similarity and cascade of turbulence. However, it is difficult to understand the self-similarity and cascade of classical turbulence in the real space because vortices are unstable and not well-defined. 

Quantum turbulence and quantized vortices exhibit advantages over classical turbulence and vortices, respectively. In superfluid $^4$He, vortices are stable topological defects and their circulation is conserved by quantization. The quantum circulation is given by $\kappa=h/m$, where $h$ and $m$ are Planck's constant and the mass of a $^4$He atom \cite{Feynman1955,Vinen1961}. Because quantum turbulence consists of well-defined elements, studies can provide a shortcut to investigate turbulence. 
The self-similarity of quantum turbulence in a wave number space such as Kolmogorov's law was studied \cite{NorePRL,NorePoF,Maurer1998,Stalp1999,Araki2002,Kobayashi2005,Kobayashi20052,Parker2005,Baggaley2012}.
However, studies on self-similarity in a real space are scarce; an example of such a study is \cite{Araki2002,Mitani2006,Kadokura2018}. We focus on the statistical laws in a real space assuming that quantum turbulence has some self-similarity.


Liquid $^4$He changes to superfluid phase at temperatures below $T_\lambda=2.17\  \mathrm{K}$, and its hydrodynamics can be described by the two-fluid model. This implies the superfluid $^4$He is a mixture of a viscous normal fluid component and an inviscid superfluid component \cite{Tisza1938,Landau1941}. The density and velocity of the superfluid component are $\rho_\mathrm{s}$, and $\bm{v}_\mathrm{s}$, respectively and those of the normal fluid component are $\rho_\mathrm{n}$, and $\bm{v}_\mathrm{n}$, respectively. The total density is $\rho=\rho_\mathrm{s}+\rho_\mathrm{n}$. The ratio $\rho_\mathrm{s}/\rho$ increases with decreasing temperature. Particularly, below approximately $1\ \mathrm{K}$, the ratio is $\rho_\mathrm{s}/\rho\simeq1$.
At finite temperatures, mutual friction acts between the two components through quantized vortices. Mutual friction can significantly shrink a vortex ring that moves in the fluid.

There are several methods to generate quantum turbulence in a superfluid $^4$He \cite{Vinen2002,Tsubota2013,Barenghi2014}, and experiments using oscillating objects have been recently conducted \cite{Jager1995,Luzyriaga1997,Yano2007,Hashimoto2007,Garg2012,Bradley2009,Bradley20092,Yano2010,Bradley2011,Bradley20123,Bradley2012,Yano2019}. A vibrating wire is a typical oscillating object. Thin wires are vibrated by a Lorentz force under a static magnetic field, which generates turbulence around the wire.
Yano $et$ $al.$ conducted a series of experiments using vibrating wires \cite{Yano2007,Goto2008,Yano2010,Yano2019}.
From the Yano group two kinds of vibrating wires, namely, a generator of turbulence and detector of vortices were discovered.
A generator wire has remnant vortices, whereas a detector wire has no remnant vortices. Although the wire velocity increases with the driving force, the two kinds of wires have different behaviors. If the driving force exceeds some critical value, the velocity of the generator wire decreases immediately and a vortex tangle is generated around it. 
However, a detector wire does not generate turbulence by itself because of the success of removing remnant vortices around it. 
If a vortex ring approaches a detector wire, it generates a vortex tangle around it using the ring as a trigger thereafter decreasing the wire velocity.
Accordingly, Yano $et$ $al.$ performed experiments using a detector wire to detect the vortex rings emitted from a vortex tangle made by a generator wire.

An important feature of the experiments is that it is possible to manage the minimum size of detectable vortex rings by altering the temperature. At $0\ \mathrm{K}$, a vortex ring moves with its self-induced velocity without shrinking.
At finite temperatures, a vortex ring shrinks in its flight and can disappear by mutual friction. The flight distance $l$ for a vortex ring with an initial radius $R_0$ disappears is given by $l = R_0/\alpha$, where $\alpha$ is the mutual friction coefficient described later. Therefore, the diameter $2R_0$ of a detectable vortex ring satisfies $2R_0 > 2\alpha D$, where $D$ is the distance between the detector and the generator wires.

Using the setup, Yano $et$ $al.$ recently observed some self-similarity of vortices emitted from a vortex tangle. This experiment discovered two important laws. First, the time of flight of vortex rings from the vortex tangle to the detector wire follows exponential distributions for any detectable minimum size. Particularly, the detection of vortex rings follows a Poisson process. This means that vortex rings are detected with frequencies depending on their sizes; hence, a vortex tangle is in a statistically steady state. Second, the vortex tangle has self-similarity. From the experiment, the relationship between the detection frequency and the minimum size of the detectable vortex rings that satisfies the power law was determined. The vortex ring size should reflect the vortex line spacings in the tangle. Therefore, the vortex tangle can have a self-similar structure in a real space.

These results show the statistical laws of a localized vortex tangle. Although there are a lot of studies on statistical laws in bulk quantum turbulence, studies in inhomogeneous or localized turbulence is scarce. 

For experiments on quantum turbulence generated by oscillating objects, several numerical simulations have been conducted. The purpose of previous simulations was to investigate the processes of growth and decay of a localized vortex tangle or the anisotropy of the emission of vortex rings from a tangle \cite{Hanninen2007,Goto2008,Fujiyama2010,Nakatsuji2014}. This study focuses on the statistical laws and the self-similarity of vortex rings emitted from a localized tangle, which differs from the previous works.
 
Using the vortex filament model, we numerically examine the dynamics and statistics of vortices emitted from a localized vortex tangle. Our goal is to examine the statistical properties of this system and to compare with the experimental results. First, we obtain a localized statistically steady vortex tangle as the source of emitted vortex rings. Second, we study the statistics of the vortex rings emitted from the tangle. In Section II, we introduce the vortex filament model and the system treated in this study. Thereafter, we describe the formation of vortex tangles in Section III.
In Section IV, we discuss statistically steady vortex tangles and introduce some theoretical concepts. Furthermore, we present the statistical laws and compare the exponents of the power laws with those of the experimental results in Section V. 
Finally, Section VI presents the conclusions.

\section{The model and system}
\subsection{Vortex filament model}
Quantized vortices in superfluid $^4$He are stable topological defects with quantized circulation and thin cores of order $1$ $\mathrm{\AA}$. Therefore, we can use the vortex filament model wherein vortices are treated as filaments. The superfluid velocity field obtained owing to quantized vortices is given by the Biot-Savart law \cite{Schwarz1985}
\begin{equation}\label{eq:BS}
\bm{v}_{\mathrm{s,BS}}(\bm{r},t)=\frac{\kappa}{4\pi}\int_L\frac{\bm{s}^\prime(\xi,t)\times(\bm{r}-\bm{s}(\xi,t))}{\mid\bm{r}-\bm{s}(\xi,t)\mid^3}d\xi \;,
\end{equation}
where $\bm{s}(\xi,t)$ denotes the position of the vortex filaments represented by the parameter $\xi$, and $\bm{s}^\prime = \frac{\partial \bm{s}}{\partial \xi}$. The integration is performed over the whole vortex filaments $L$.
At finite temperatures, mutual friction affects the motion of vortices. If there are neither boundaries nor applied superfluid flow, the equation of motion becomes
\begin{equation}\label{eq:EoM}
\frac{d\bm{s}}{dt}=\bm{v}_{s,BS}+\alpha\bm{s}\times\left(\bm{v}_n-\bm{v}_{s,BS}\right)-\alpha^\prime\bm{s}^\prime\times\left[\bm{s}^\prime\times\left(\bm{v}_n-\bm{v}_{s,BS}\right)\right],
\end{equation}
where $\alpha$ and $\alpha^\prime$ are the coefficients of friction depending on the temperature. In particular, $\alpha,\alpha^\prime=0$ at $T=0\ \mathrm{K}$.

The vortex lines are discretized into a number of points held at a minimum space resolution $\Delta \xi = 0.8\ \mathrm{\mu m}$. The integration in time is performed using the fourth-order Runge-Kutta scheme, wherein the time resolution is $\Delta t=10\  \mathrm{\mu s}$. We use the traditional method to artificially reconnect two vortices that approach each other within $\Delta\xi$ \cite{Adachi2010}. We delete the vortex rings whose lengths are shorter than $6\Delta\xi$.

Such reconnection of quantized vortices can be related to the dissipation mechanism of quantum turbulence at very low temperatures with negligible mutual friction. The numerical simulation of the Gross-Pitaevskii model shows that reconnections emit phonons of short wavelengths comparable to the coherence lengths and causes the dissipation \cite{Leadbeater2001}. However, the vortex filament model cannot describe the phonon emission. The change of vortex length at each artificial reconnection is negligible compared to the vortex dynamics in the large scales. This study focuses on the statistical laws at large scales wherein details of each reconnection is not relevant.
\subsection{The system}
The motivation of this study is to reproduce the statistical laws observed by the experiment and reveal the self-similarity of the system. To achieve this, we first obtain a localized stationary vortex tangle as the source and thereafter observe the emission of vortex rings from the tangle.

\begin{figure}
\includegraphics[scale=0.2]{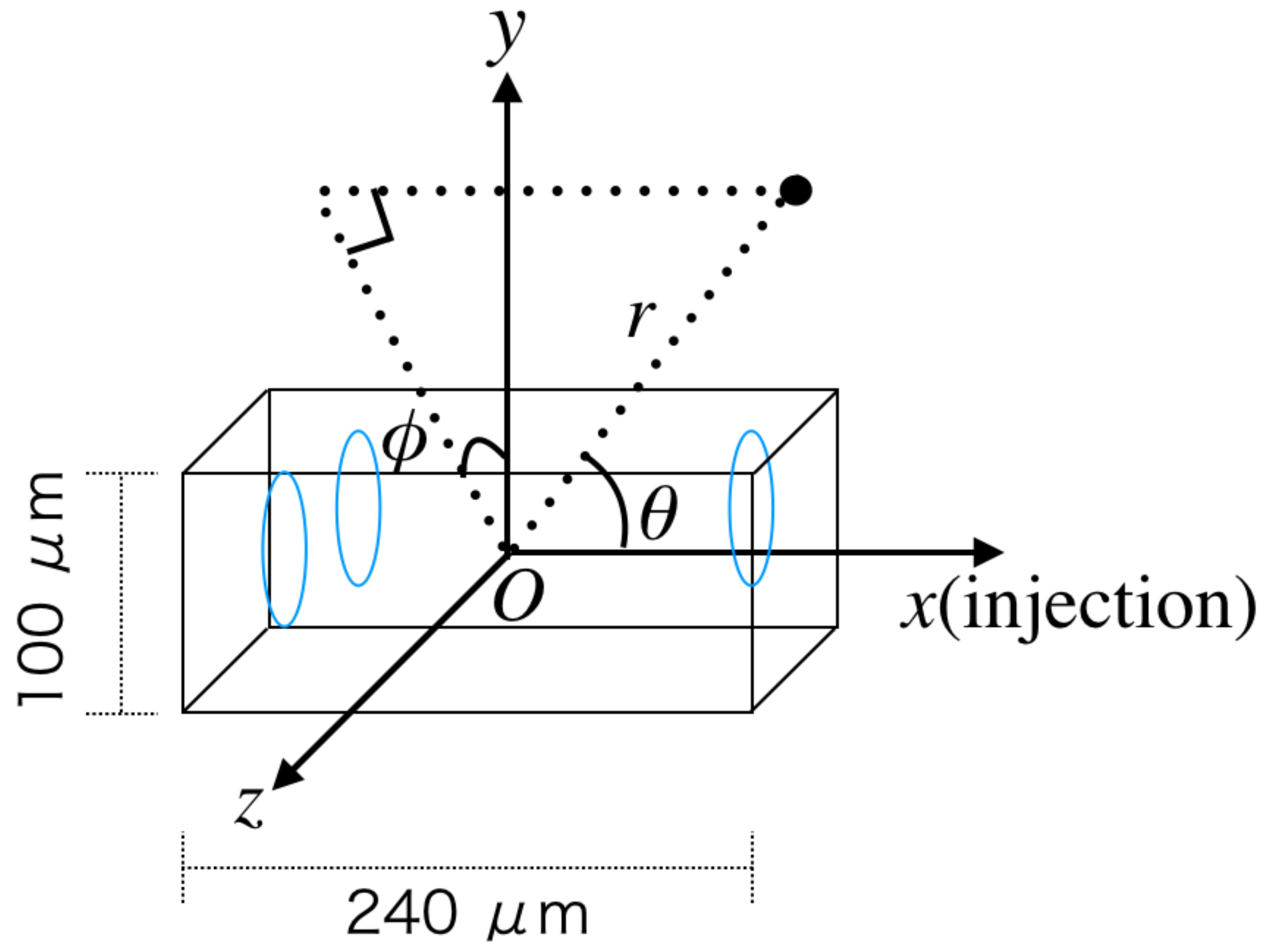}
\caption{\label{fig:axis} The coordinate system of this system. We set the $x$-axis as the injection direction of vortex rings. Vortex rings are injected from two parallel $100\  \mathrm{\mu m}\times100\ \mathrm{\mu m}$ square vortex sources at a fixed frequency.}
\end{figure}

The method of generating a localized vortex tangle is a key problem in our simulation.
We are predominantly interested in the emission of vortex rings from a localized vortex tangle.
We use a novel method that differs from those in previous simulations \cite{Goto2008,Hanninen2007,Fujiyama2010,Nakatsuji2014} to generate a dense localized vortex tangle that can emit many vortex rings.
As shown in Fig. \ref{fig:axis}, we prepare two parallel $100\ \mathrm{\mu m} \times 100\  \mathrm{\mu m}$ square vortex sources that inject vortex rings of some size at a fixed frequency from random positions in each square.
 The distance between the two parallel sources is $240\ \mathrm{\mu m}$. The parameters of this simulation are the injection frequency $f$ and the diameter $2R_0$ of the injected vortex rings. Now, let $f$ be of order $1\ \mathrm{kHz}$ corresponding to the frequency of the vibrating wire and $2R_0$ be of order $10\ \mathrm{\mu m}$ corresponding the amplitude of the vibration \cite{Yano2019}. 
 To be later described in detail, we maintain injecting vortices from the two sources and generating a localized vortex tangle.
 These tangles expand orthogonally to the direction of the injection, as shown in Fig. \ref{fig:tangle}. These tangles are regarded as source vortex tangles formed from a vibrating object. 

Furthermore, we simulate the detection of vortices by the detector. One detector was used in the experiment. The experiment was repeated severally to use the observations to obtain the statistical law \cite{Yano2019}.
However, our simulation allocates many detectors around the tangle.
 The vortex tangle and the emitted vortex rings should be symmetric about the azimuthal angle $\phi$.
The vortex rings are emitted orthogonally to the direction of injection, as described in the next section.
We collect the data on the vortex rings emitted from a vortex tangle at a fixed distance of $400\ \mathrm{\mu m}$ from the origin and eliminate from the vortices we follow. 

\section{properties of the localized vortex tangle}
Comparing to the experiments \cite{Yano2019}, the success of our simulation depends on obtaining statistically steady localized vortex tangles by the method described in the last section.
In this section, we describe the development of the vortex tangle and show that a statistically steady vortex tangle can be generated. Thereafter, we describe the statistics of the observations of the vortex rings emitted from the vortex tangle. 

\subsection{Development of vortex tangle}
\begin{figure}[h]
\includegraphics[scale=0.24]{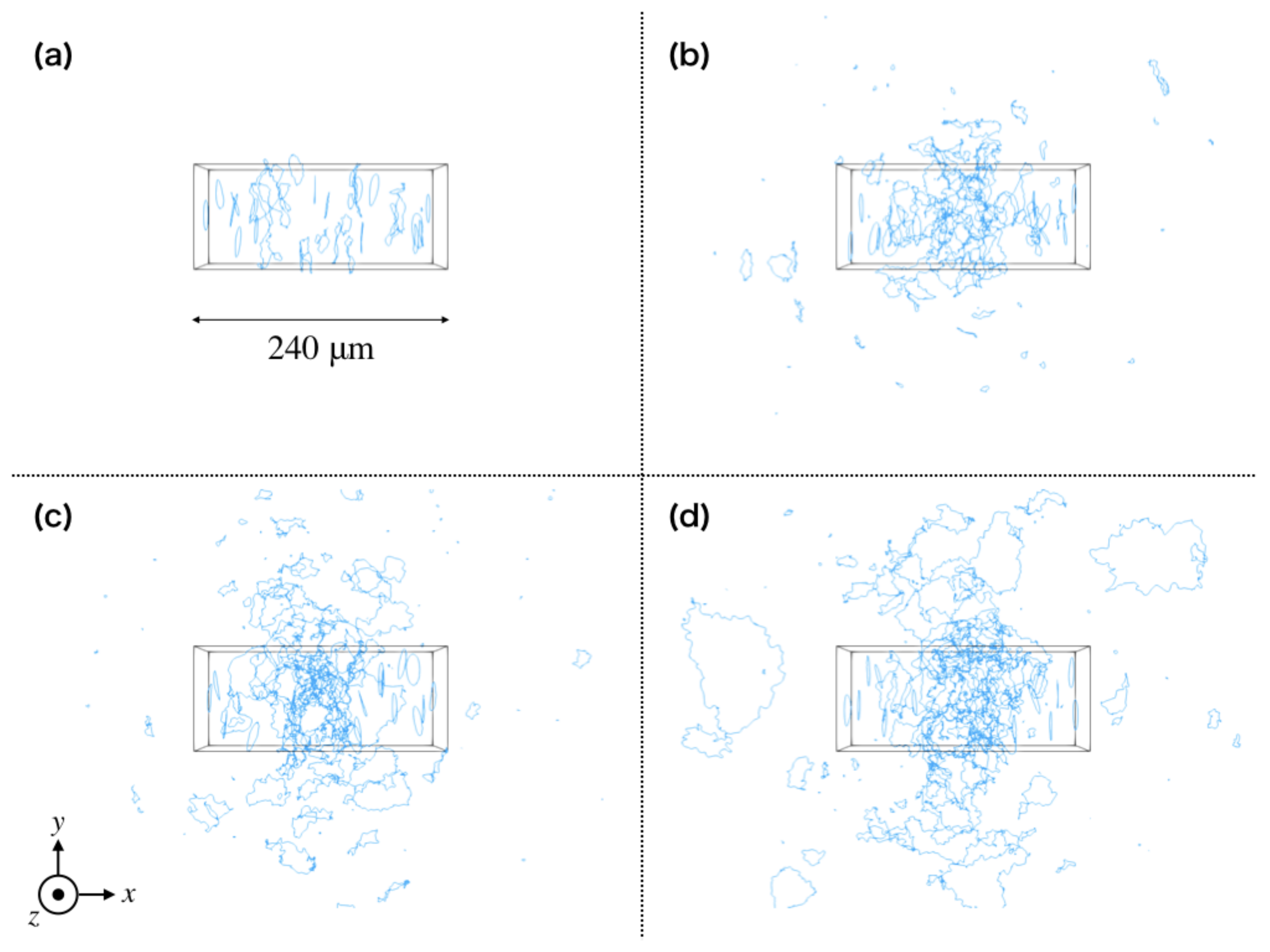} 
\caption{\label{fig:tangle} Development of vortex tangle in $f=1000\ \mathrm{Hz}$ and $2R_0=30\ \mathrm{\mu m}$ at time (a)$t=0.02$ $\mathrm{s}$ (b)$t=0.06$ $\mathrm{s}$ (c)$t=0.16$ $\mathrm{s}$ (d)$t=0.40$ $\mathrm{s}$, respectively. The black rectangular box refers to the box in Fig. \ref{fig:axis}. Although the vortex tangles of (a) and (b) grow, those of (c) and (d) are saturated inside the box.}
\end{figure}

Figure \ref{fig:tangle} shows a typical development of a vortex tangle. If vortex sources begin to inject vortex rings, they form a small nucleus of a vortex tangle around the origin if they frequently collide (See Supplemental Material). Thereafter, the nucleus absorbs subsequent vortex rings and develops a vortex tangle. 
Although this explanation is satisfactory, the statistical steadiness or unsteadiness of the resulting localized vortex tangle is nontrivial. 

\begin{figure}[h]
\includegraphics[scale=0.4]{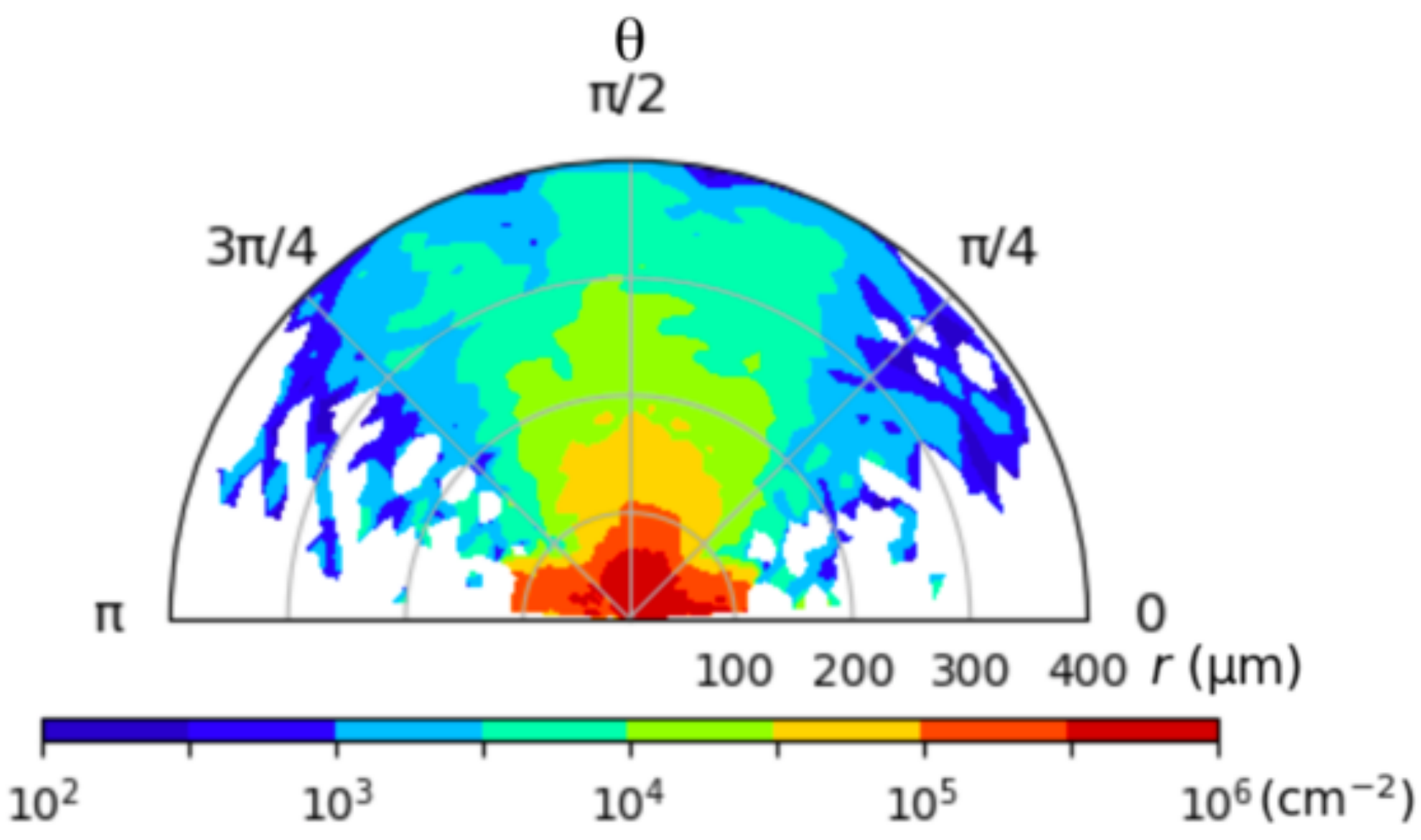} 
\caption{\label{fig:densdevelop} The time averaged($t=0.4$ - $0.6\ \mathrm{s}$) distribution of the vortex line density ($\mathrm{{cm}^{-2}}$) of the vortex tangle in a $r(\mathrm{\mu m})-\theta (\mathrm{rad})$ plane in the log scale. The condition is $f=1000\ \mathrm{Hz}$ and $2R_0=30\ \mathrm{\mu m}$. Because the vortex tangle is symmetric around the azimuthal angle $\phi$, the distribution is obtained by integrating over $\phi$.}
\end{figure}
Figure \ref{fig:densdevelop} shows the vortex line density distribution after the tangle develops significantly and is statistically steady. The density decreases with increasing distance from the origin. The density is concentrated around $\theta=\frac{\pi}{2}$ because of the symmetry of the system.
\begin{figure}[h]
\includegraphics[scale=0.21]{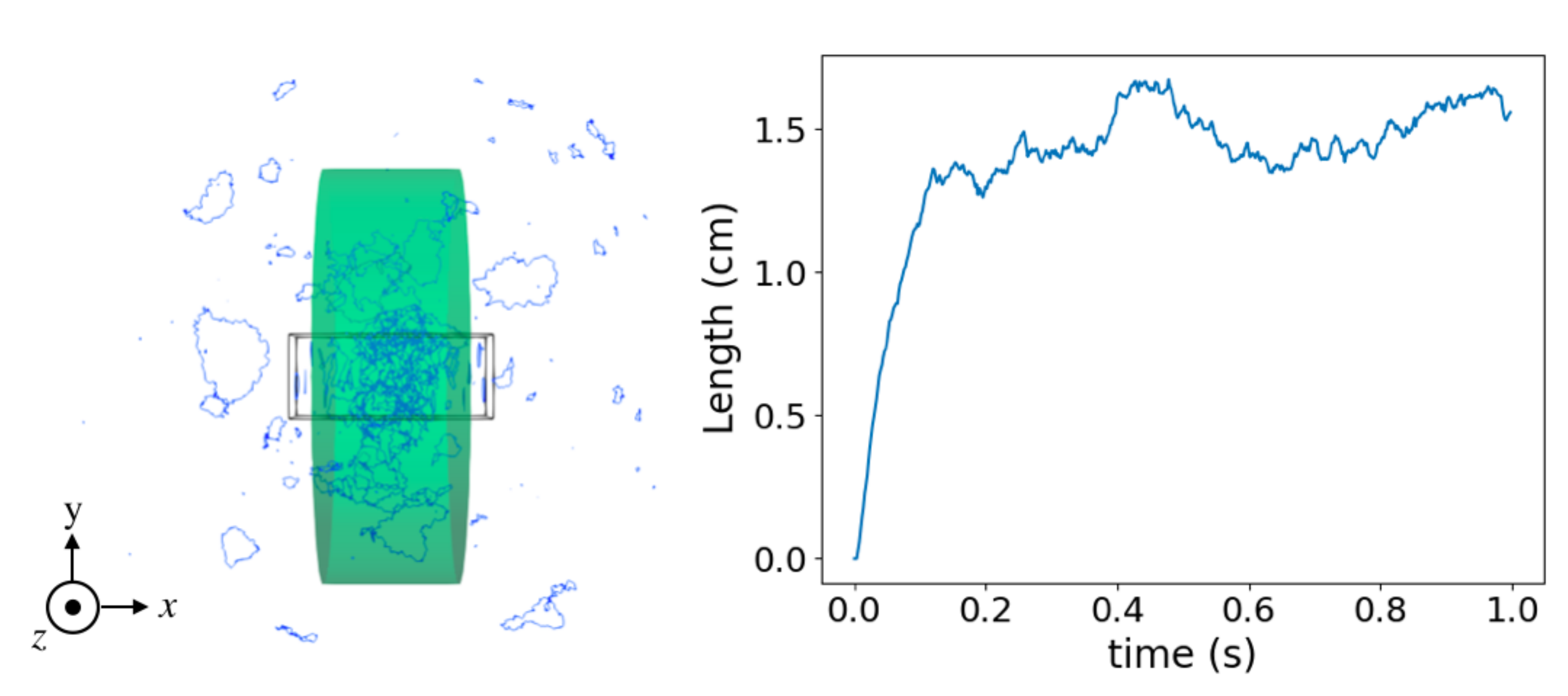} 
\caption{\label{fig:developtime} The vortex line length in the cylindrical volume in $2R_0=30$ $\mathrm{\mu m}$ and $f=1000$ $\mathrm{Hz}$. The cylindrical volume has its height 160 $\mathrm{\mu m}$ and its radius 250 $\mathrm{\mu m}$. This cylinder covers the vortex tangle. The black box in the left figure is as same as that in Fig. \ref{fig:tangle}. The figure on the right shows development of the vortex line length. The vortex tangle becomes statistically steady after approximately $t=0.1$ $\mathrm{s}$.}
\end{figure}

Thereafter, we directly investigate the properties of the tangle. The vortex distribution in Fig. \ref{fig:densdevelop} includes emitted vortex rings and a localized tangle.
From Fig. \ref{fig:densdevelop}, we know that the tangle expands orthogonally to the $x$--axis and can estimate the approximate size of the tangle.
We assume a cylindrical volume with height $160\ \mathrm{\mu m}$ and radius $250\  \mathrm{\mu m}$ that covers the vortex tangle and reflects its symmetry. The centroid of the volume is placed at the origin, and the bottom is orthogonal to the $x$--axis. Figure \ref{fig:developtime} shows the development of the total vortex line length in the cylindrical volume. The vortex line length increases with time and is statistically steady after approximately $t=0.1\ \mathrm{s}$. 

\begin{figure}[h]
\includegraphics[scale=0.24]{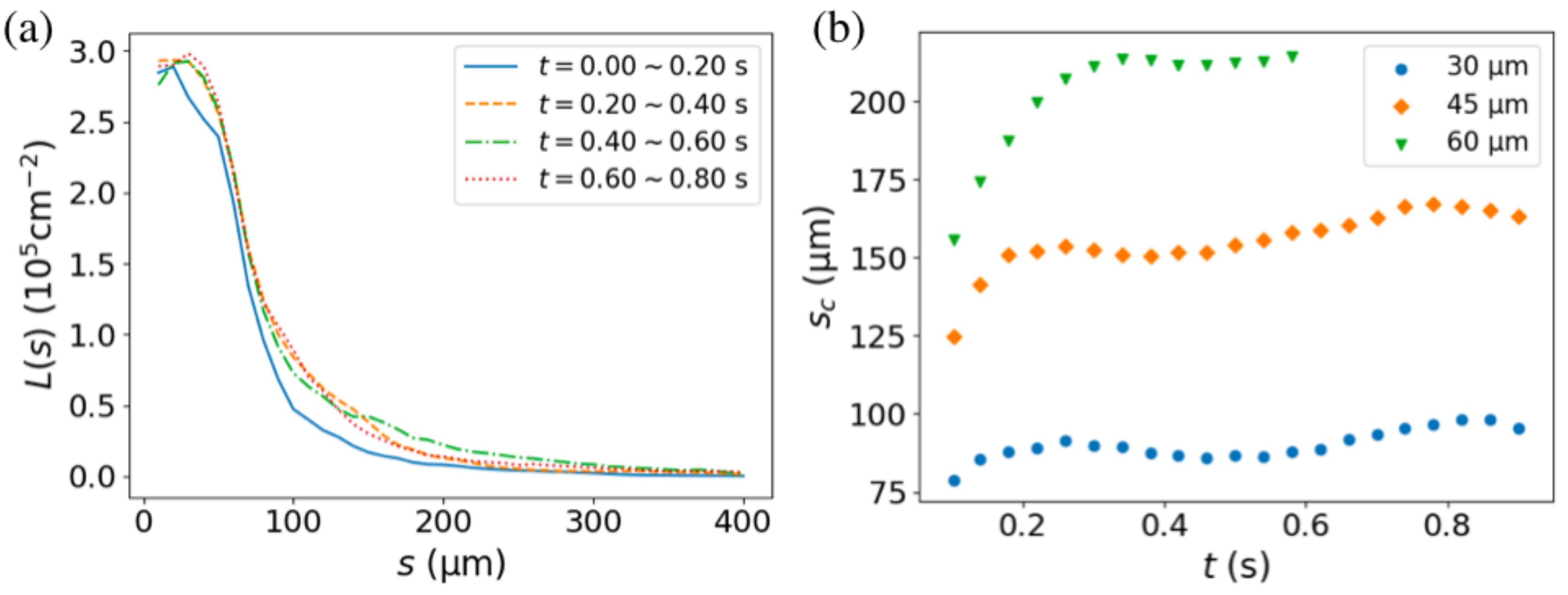}
\caption{\label{fig:dens}(a) The time averaged vortex line density distribution in $f=1000\ \mathrm{Hz}$ and $2R_0=30\ \mathrm{\mu m}$. The colors represent the distribution averaged over the different time intervals. (b) The time development of $s_c$ in $f=1000\ \mathrm{Hz}$. The colors represent the different injected vortex ring sizes $2R_0$. The density is averaged for the time interval between $t-0.1\ \mathrm{s}$ and $t+0.1\ \mathrm{s}$.}
\end{figure}
To characterize the steady states, we investigate the distribution $L(s)$ of the vortex line density in the hollow cylindrical volumes with height, inner radius, and outer radius as $160\ \mathrm{\mu m}$, $s-ds$, and $s$, respectively.
The centroid of the volume is placed at the origin, and the bottom is orthogonal to the $x$--axis.
Fig. \ref{fig:dens}(a) shows the time-averaged distribution $L(s)$. Because the distribution does not change significantly after approximately $t=0.2\ \mathrm{s}$, the tangle is observed to be statistically steady about the vortex distribution and its total length.

From the distribution of the vortices, we estimate the size of the tangle, although there is some arbitrariness for defining the size of the tangle.
We define the tangle size $s_c$ such that the density $L(s)$ decreases to $10^{5}\  \mathrm{cm}^{-2}$ in the volume.

Figure \ref{fig:dens}(b) shows the time development of $s_c$ for $2R_0=30,\ 45,\ 60$ $\mathrm{\mu m}$ and $f=1000\ \mathrm{Hz}$. The tangle size converge to a finite value in each condition. 
The size of tangle in the statistically steady state increases with $2R_0$.

The subsequent steadiness or unsteadiness of a developed vortex tangle is nontrivial.
Injected vortex rings are shuffled to form a localized vortex tangle. The vortex tangle emits vortex rings that operate as the dissipation for the tangle.
The statistically steady state is sustained by the equilibrium of the vortex ring injection, the deletion of small rings, and the vortex ring emission from the tangle.
We do not know if such statistical steady states are consistently obtained for arbitrary values of $2R_0$ and $f$. This can be investigated in future studies. 

\subsection{The vortex line length of vortex tangle }
If the injection frequency $f$ and the size $2R_0$ of the injected vortex rings are reduced, a statistically steady vortex tangle may not be generated because vortex rings do not frequently collide and generate no nucleus of a tangle.
\begin{figure}[h]
\includegraphics[scale=0.45]{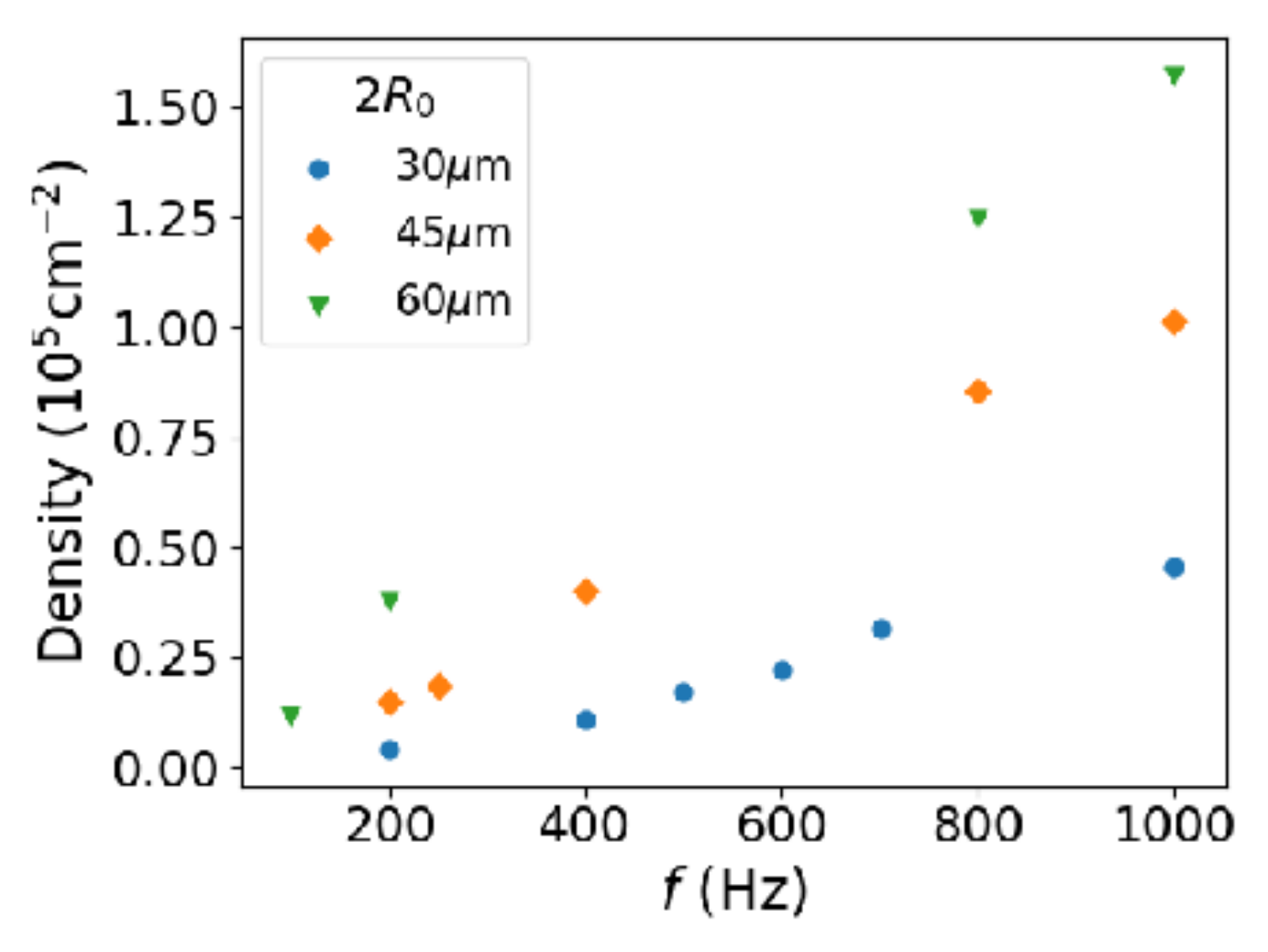}
\caption{\label{fig:souzu2} The mean vortex line density in the cylindrical volume. The horizontal axis is the frequency $f$.}
\end{figure}

We calculate the dynamics with varying $f$ and $2R_0$ to study the conditions for the generation of a statistically steady vortex tangle. 
Figure \ref{fig:souzu2} shows that the statistically steady vortex line density in the cylinder increases with $f$ and $2R_0$.
\begin{figure}[h]
\includegraphics[scale=0.45]{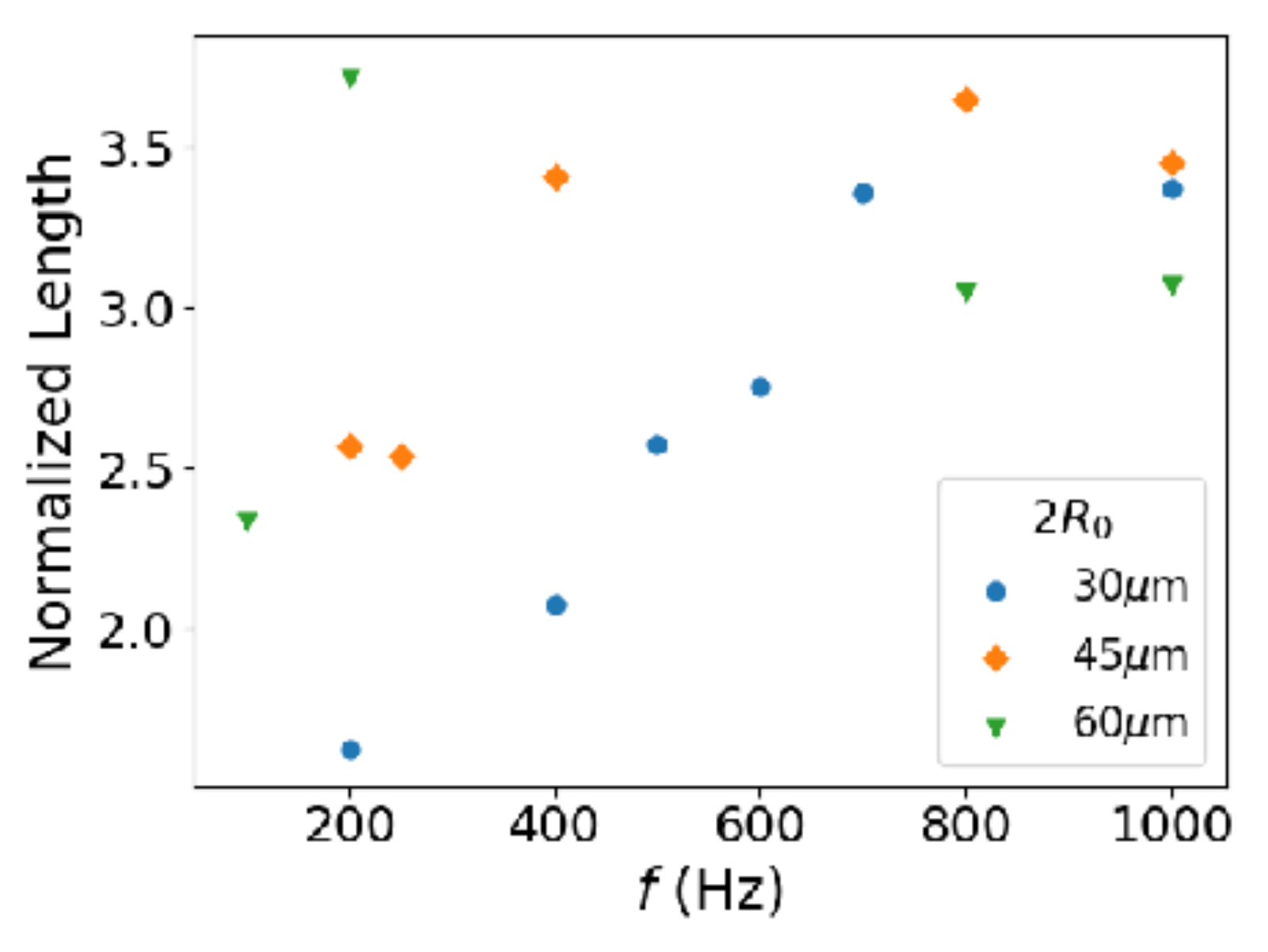}
\caption{\label{fig:normed} The vortex line length normalized by $L_\mathrm{n}$ of Eq. (\ref{noml}).
Here $2R_0$ and $l$ are the size of injected vortex rings and the height of the cylindrical volume, respectively.}
\end{figure}

The appearance (or no appearance) of a nucleus of a vortex tangle determines its growth. If no nucleus is formed, no vortex tangle occurs. Investigating the conditions for the formation of a nucleus aids the determination of the characteristic vortex lengths. If injected vortex rings collide and interact, the vortex length increases. We consider the vortex length when counter-propagating rings pass through and no nucleus is formed. This consideration yields the characteristic vortex length used to normalize the vortex line length in the cylinder. The characteristic length can be determined from the geometry of Fig. 1. If counter-propagating vortex rings never collide, we can obtain the time $4\pi lR_0/(\kappa\log(R_0/r_\mathrm{c}))$ that taken for an injected ring to pass through the cylindrical volume because of the self-induced velocity of the vortex ring $v\sim\kappa/4\pi R_0\times\log(R_0/r_\mathrm{c})$. Here, $l$ is the height of the cylindrical volume. Because the length of an injected vortex ring is $2\pi R_0$, the total vortex line length $L_\mathrm{n}$ required is 
\begin{align}\label{noml}
L_\mathrm{n}&=2\times f\times \frac{4\pi lR_0}{\kappa\log(R_0/r_\mathrm{c})}\times 2\pi R_0 \notag \\ 
&=\frac{16\pi^2lfR_0^2}{\kappa\log(R_0/r_\mathrm{c})}.
\end{align}
The vortex line length normalized by $L_n$ is shown in Fig. \ref{fig:normed}. 
This quantity is the amplification factor of the vortex line length. 
If the normalized length exceeds unity, the mere group of ballistic vortex rings develops to a vortex tangle.
Increasing the injected vortex ring size and the frequency of the injection increases the length that converges to approximately $3.5$ independently of $2R_0$. This means that the vortex line length is proportional to $fR_0^2$.

\subsection{Emission of vortex rings from a vortex tangle}
\begin{figure}[h]
\includegraphics[scale=0.45]{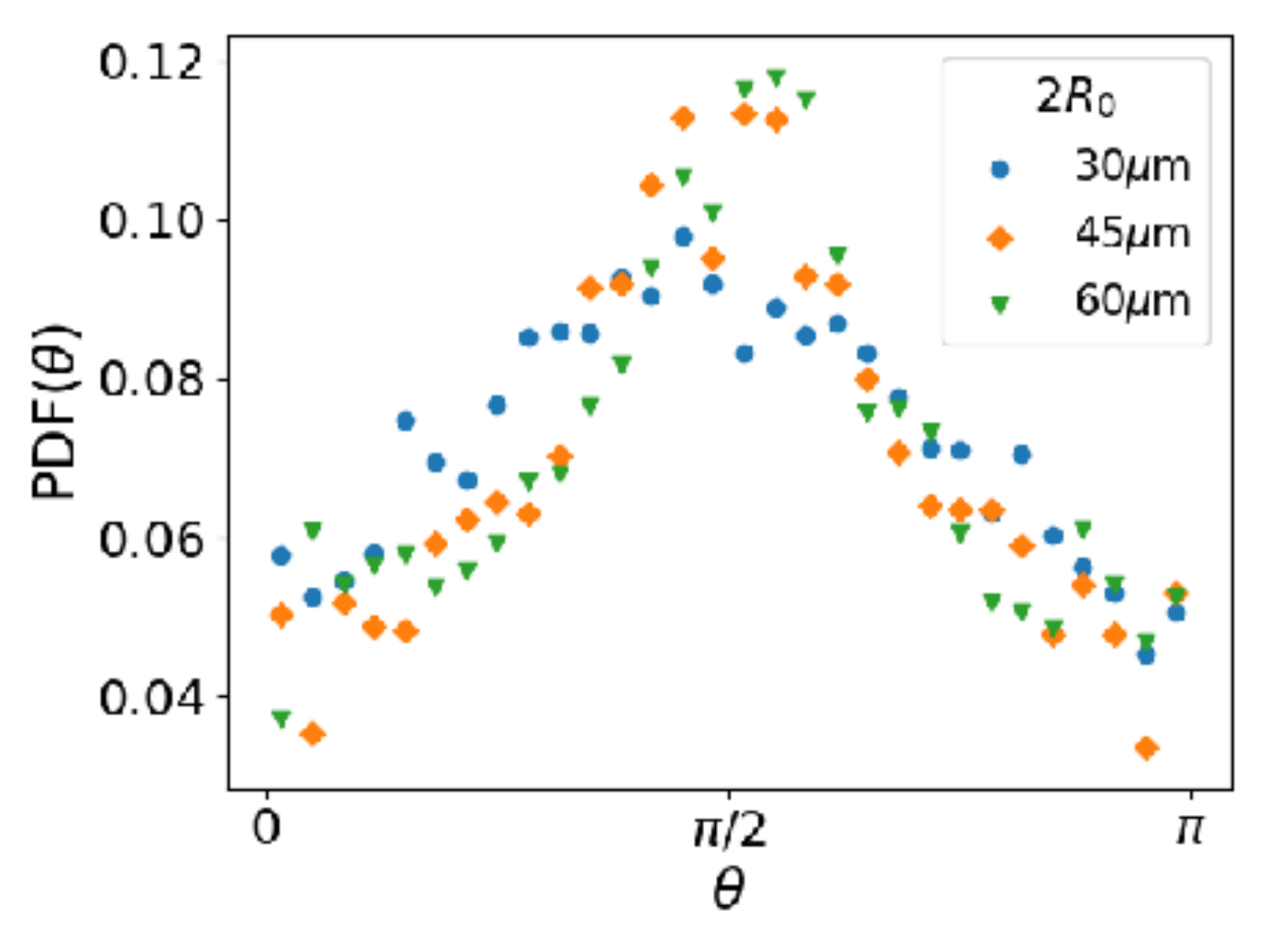}
\caption{\label{fig:pt} The probability density function, PDF $(\theta)$, of the direction of the emitted vortex rings about $\theta$. The PDF is obtained from the number of vortex rings received by the detectors within $2\pi\sin\theta d\theta$. }
\end{figure}
Although the distribution of the emitted vortex rings is isotropic about $\phi$, that in the direction $\theta$ is anisotropic.
The probability density function (PDF) in the case of the direction $\theta$ of the vortex rings emitted from the vortex tangle is shown in Fig. \ref{fig:pt}.
The data of the vortex rings are collected by the detectors placed at $400\ \mathrm{\mu m}$ from the origin.
 The emission of the vortex rings is concentrated around $\theta=\frac{\pi}{2}$. 

\section{Statistically steady tangle}

A statistically steady vortex tangle should emit vortex rings of each size with the corresponding frequency. Particularly, the emission frequency of some size is governed by a function $f(R)$ that depends only on the vortex ring size $R$. This statistically steady concept is essential in this study. From our simulation, $f(R)$ is a power of $R$, that is, the emission of the vortex rings from a tangle has some self-similarity. We introduce some theoretical concepts of this self-similarity in this section.

\subsection{Poisson process}
To investigate the statistics of emitted vortex rings, the experiment is conducted assuming a Poisson process \cite{Yano2019}.
The Poisson process is a stochastic process based on an exponential distribution. First, we describe the derivation of the process considering the conditions of the experiment.
We make the following assumptions. A detector catches $n$ vortex rings in the time interval $[0, T]$ that is partitioned into smaller intervals $\Delta t$. Thus, the probability that a vortex ring is detected in $\Delta t$ is $\Delta t\frac{n}{T}$. This is based on the assumption that $\Delta t$ is significantly small such that the number of vortex rings received in $\Delta t$ is $0$ or $1$. The probability that a vortex ring is not detected in $[0,t]$ and detected in [$t$,$t+\Delta t$] is
\begin{equation}\label{dtprob}
\left(1-\Delta t\frac{n}{T}\right)^{\frac{t}{\Delta t}}\times\Delta t\frac{n}{T}\;.
\end{equation}
If $P(t)$ is the probability that a vortex ring is detected in $[0,t]$, we have
\begin{equation}\label{eq:P}
P(t+\Delta t)-P(t)=\left(1-\Delta t\frac{n}{T}\right)^{\frac{t}{\Delta t}}\times\Delta t\frac{n}{T}\;.
\end{equation}
The PDF $F(t)$ is defined by the probability $F(t)dt$ that a vortex ring is detected in [$t$,$t+\Delta t$]. Hence, $F(t)$ is given by
\begin{align}\label{eq:DOP}
F(t)=\lim_{\Delta t \rightarrow 0}\frac{P(t+\Delta t)-P(t)}{\Delta t}&=\frac{1}{t_1}\exp\left( -\frac{t}{t_1}\right)
\end{align}
where $t_1\equiv\frac{T}{n}$ is the mean detection interval. 
Finally, integrating $F(t)$ yields
\begin{equation}
P(t)=\int_0^t F(t^\prime)dt^\prime=-\exp\left(-\frac{t}{t_1}\right)+1 
\end{equation}
and
\begin{equation}\label{eq:poason}
1-P(t)=\exp\left(-\frac{t}{t_1}\right).
\end{equation}
This indicates a Poisson process. If this relation is confirmed, the interval $t_1$ will be constant indicating that the vortex tangle is statistically steady and emits vortex rings continuously and steadily. Yano $et$ $al.$ observed this relation that indicates that the generator wire generates a statistically steady vortex tangle \cite{Yano2019}. This is a motivation for our investigation to obtain a statistically steady vortex tangle in the present simulation.

\subsection{The self-similarity}
Suppose that the vortex tangle has self-similarity in real space. We find the power law between the emission frequency and vortex ring size. The power law was deduced from experiments in \cite{Yano2019}. This self-similarity is understood from the following discussions. We define the number $n(x,t)dx$ of vortex rings with diameter in $[x,x+dx]$ emitted from a tangle in $[0,t]$. The number $N_{2R>2R_C}(t)$ of vortex rings with diameters larger than $2R_C$ emitted in $[0,t]$ is $N_{2R>2R_C}(t)=\int^\infty_{2R_C}n(x,t)dx$. 
Because the vibrating
wire experiments observe only vortex rings larger than
some minimum size, we also consider the number of vortex rings larger than some minimum size.
 When the vortex tangle is statistically steady, the number of emitted rings $n(x,t)dx$ is shown by $g(x)tdx$ and the vortex tangle emits vortex rings with various sizes.
 The $g(x)dx$ is the emission frequency of vortex rings with sizes in $[x,x+dx]$. Therefore, the frequency $f_{2R>2R_C}$ of the emission of rings larger than ${2R_C}$ is given by
\begin{equation}\label{eq:2rc}
f_{2R>2R_C}=\int^\infty_{2R_C}g(x)dx.
\end{equation}
The distribution of vortices in the tangle should be determined by that of the emitted vortex rings.
Several literature have reported that the size distribution of vortex rings in a tangle, that shows that the number of vortex rings decreases with ring sizes by some power law \cite{Araki2002,Mitani2006}. 
Assuming the emission of the vortex rings is self-similar, the frequency $g(x)$ can be written as $g(x)=x^{-\alpha}$ such that 
\begin{align}
f_{2R>2R_C}&=\int^\infty_{2R_C}x^{-\alpha}dx \propto (2R_C)^{-\alpha+1}.
\end{align}
Thus, the power law comes from the self-similarity of the size distribution of the vortex rings in the tangle.

\section{Statistical laws}
We describe the statistical laws of the vortex rings emitted from a vortex tangle. The first law is the Poisson process of the detection of the vortex rings emitted from a tangle. The second is the power law between the frequency of the emission and the vortex ring size. 

The statistically steady vortex tangle emits vortex rings, as mentioned in Section IV.
In this section, we numerically investigate the probability that detectors receive vortex rings with diameters larger than some minimum diameter $2R$. Experiments are performed using one detector. The experiment is repeated to determine the statistics. However, our simulation involved one simulation with $2000$ detectors.

\subsection{Poisson process}
\begin{figure}
\includegraphics[scale=0.4]{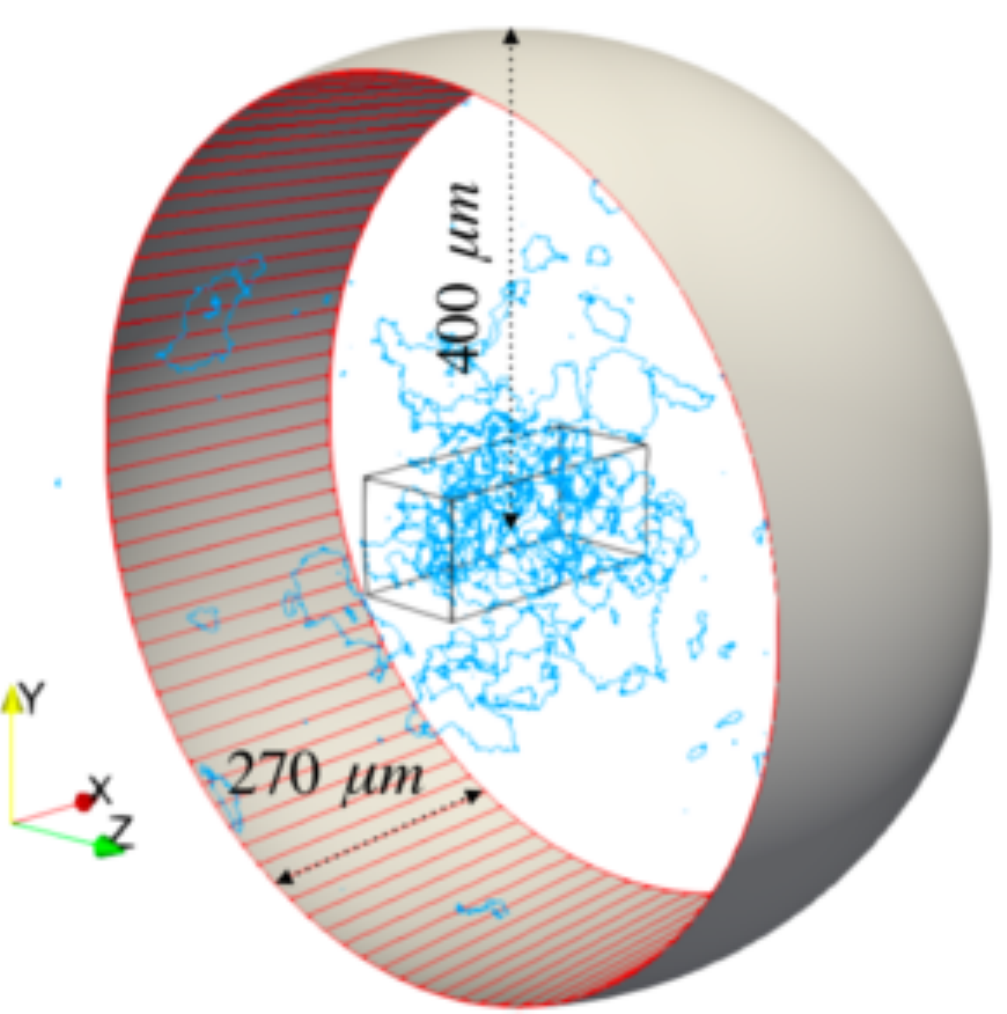}
\caption{\label{fig:detectors} The schematic figure of the arrangement of the detectors around the vortex tangle. The detectors are arranged symmetrically with the width 270 $\mathrm{\mu m}$}
\end{figure}
We position detectors at a fixed distance of $400\ \mathrm{\mu m}$ from the origin and orthogonal to the $x$-axis as shown in Fig. \ref{fig:detectors}. In this study, the number of detectors $N_\mathrm{det}$ is $2000$.
The simulation indicates the number $N(t)$ of detectors that receive at least one vortex ring in $[0,t]$ \cite{memo}. The probability $P(t)$ is given by 
\begin{equation}\label{eq:detector}
P(t)=N(t)/N_{\mathrm{det}}.
\end{equation}
This relation is a key idea that relates our simulation to the experiment.
\begin{figure}[h]
\includegraphics[scale=0.45]{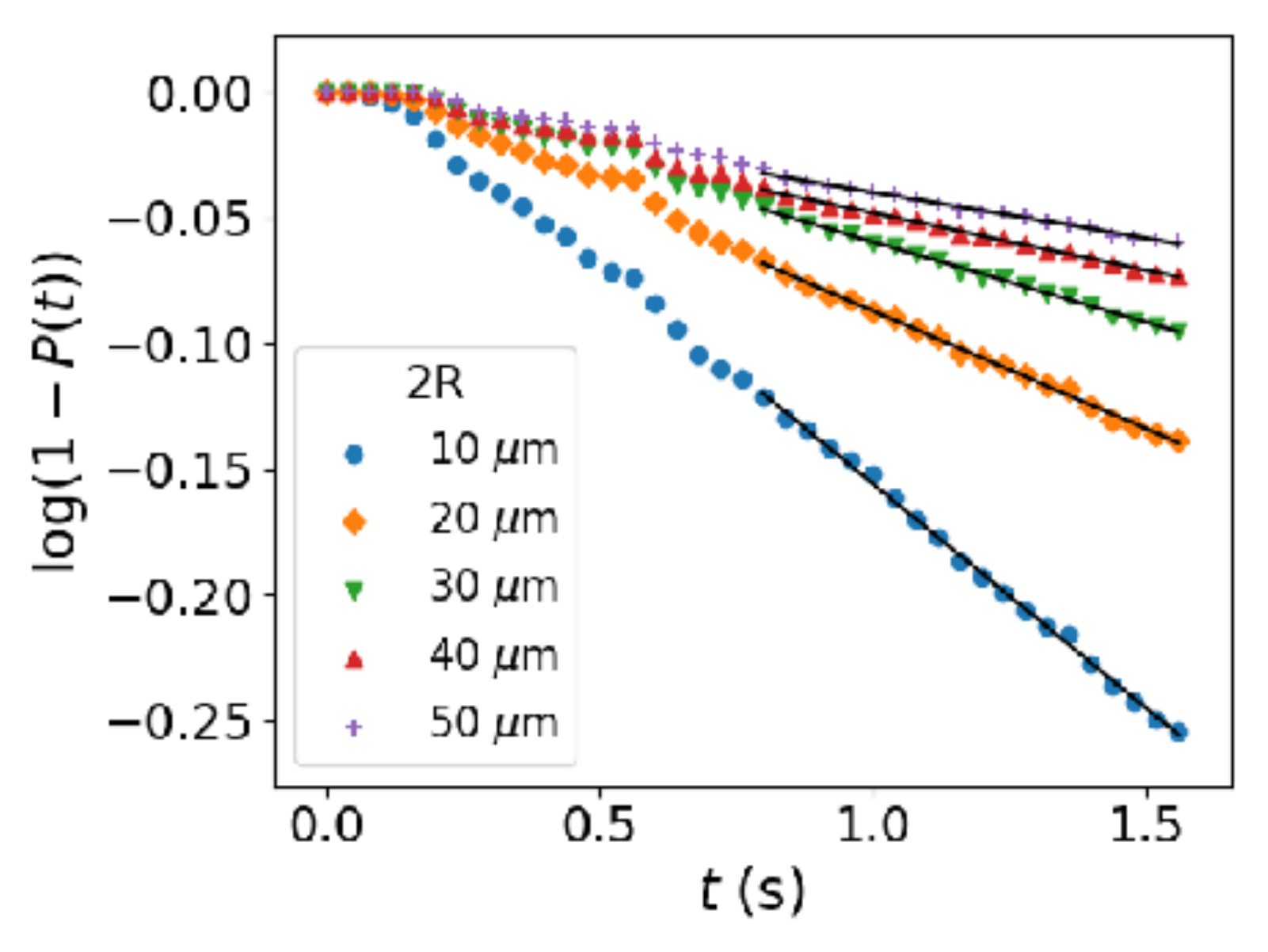} 
\caption{\label{fig:poisson} 
The time development of $1-P(t)$ in $f=1000$ $\mathrm{Hz}$ and $2R_0=30$ $\mathrm{\mu m}$. Here, $2R$ refers to the minimum size of detectable vortex rings.}
\end{figure}

Figure \ref{fig:poisson} shows the results of the simulation with $f=1000\ \mathrm{Hz}$ and $2R_0=30\ \mathrm{\mu m}$. They satisfy Eq. (\ref{eq:poason}); hence, our simulation reproduces the Poisson process observed experimentally. These slopes indicate the detection frequencies $t_1^{-1}$. 

\begin{figure}[h]
\includegraphics[scale=0.45]{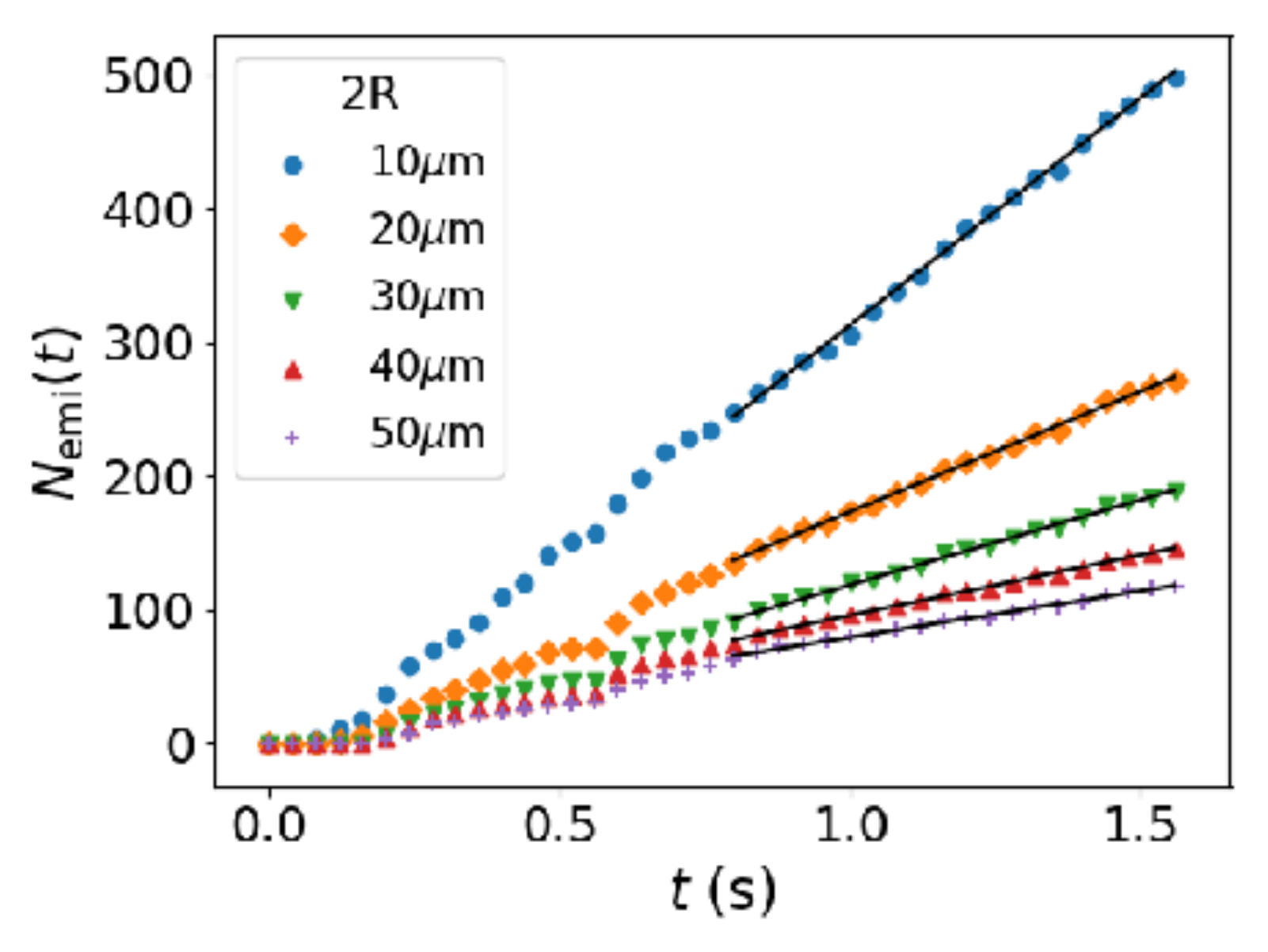}
\caption{\label{fig:kari}The number $N(t)$ of the vortex rings emitted in [0,t]. The frequency of emission converges at finite number in each minimum size. Here, $2R$ refers to the minimum size of detectable vortex rings.}
\end{figure}

To directly investigate the property of a vortex tangle, we examine the emission frequency. Figure \ref{fig:kari} shows the number $N_{\mathrm{emi}}(t)$ of vortex rings that have diameter larger than $2R$ and emitted in $[0,t]$. We can confirm the linear relationship $N_{\mathrm{emi}}(t)=t/t_1$ with the emission frequency $t_1^{-1}$. This figure shows that the frequency becomes constant indicating that the vortex tangle becomes statistically steady.

\subsection{The power law}

\begin{figure}[h]
\includegraphics[scale=0.45]{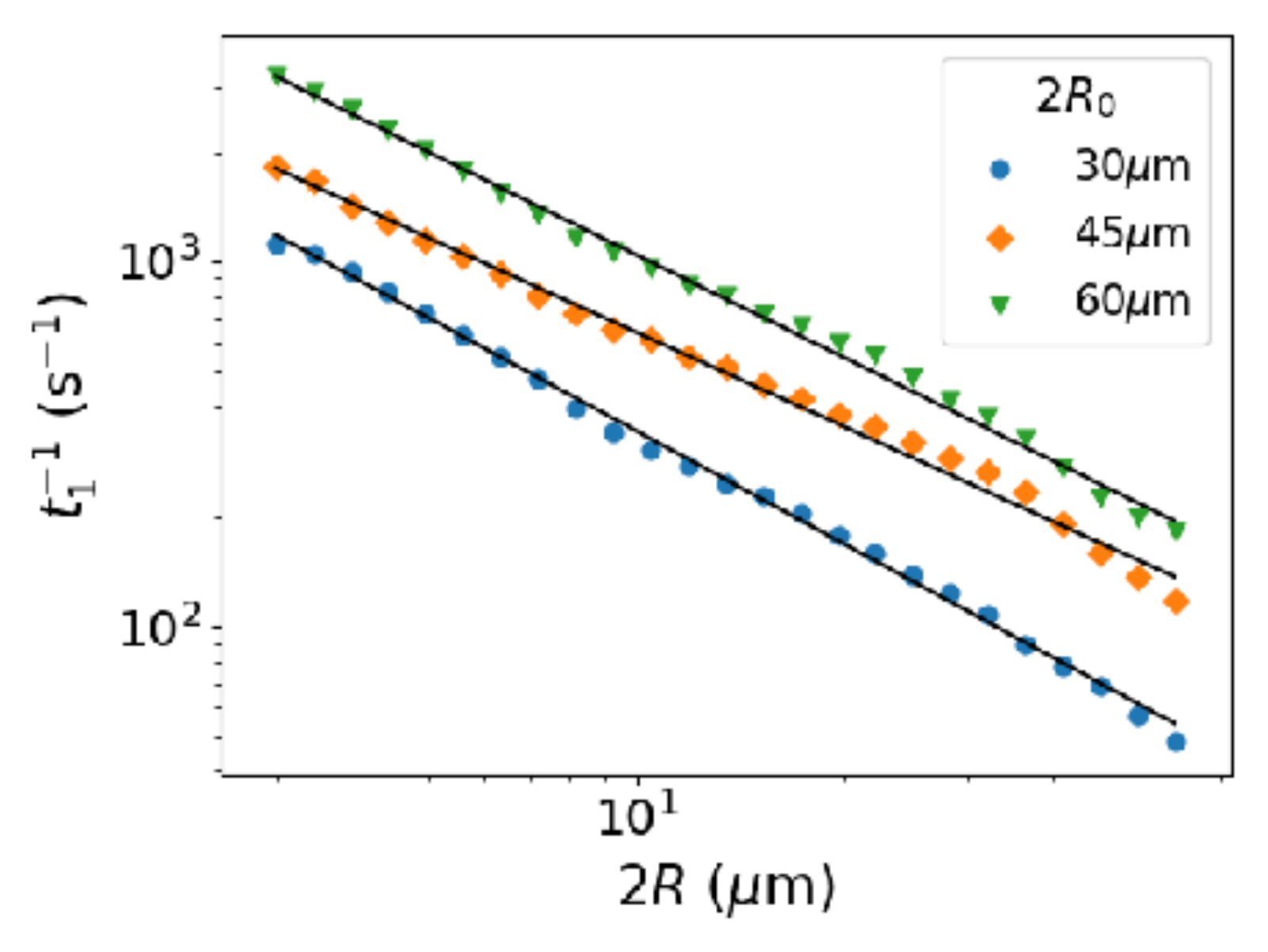} 
\caption{\label{fig:power law} The relationship between $t_1^{-1}$ and $2R$ in the log-log scale. }
\end{figure}

The power law between $t_1^{-1}$ and $2R$ indicates the self-similarity of the vortex rings emitted from the localized vortex tangle. From Fig. \ref{fig:power law}, the emission frequency $t_1^{-1}$ satisfies the power law of $2R$ for three different values of $2R_0$.
For $f=1000\ \mathrm{Hz}$ and $2R_0=30\ \mathrm{\mu m}$, the power law $t_1^{-1}=(2R)^{-1.03\pm 0.01}$ is obtained by the least squares method. Therefore, we can obtain results similar to the experiments. Here, we determine the slope in the range up to $60\ \mathrm{\mu m}$. 
 They show power laws, however, they deviate for $2R>10\ \mathrm{\mu m}$.
We propose two reasons for the deviation.
The first may come from the rare events catching such large vortices. Second, there may be a difference between the emission mechanism of vortex rings smaller and larger than the size of the tangle. The large rings can be emitted only from the surface of the tangle, whereas the small ones can be emitted from the surface or inside the tangle.
This result shows that the distribution of vortex rings emitted from the localized vortex tangle has self-similarity, that may reflect the self-similarity of the vortex size distribution of a vortex tangle.

Finally, we compare the power exponent obtained from this study with that obtained from the experiments. 
In the experiments, the exponent depends on the turbulence generation power. Yano $et$ $al.$ observed that the exponents increased to $-2.5$, $-1.6$, and $-1.5$ as the power increased to $40\ \mathrm{pW}, 150\ \mathrm{pW}$ and $1000\ \mathrm{pW}$, respectively. 
Whereas our simulation shows power exponents $-0.86\sim-1.03$, that differ from the experimental results because the vortices become significantly dense to be calculated numerically. 

The difference in the exponents between the simulation and the experiments may connected with the emission power. The energy $\epsilon$ of the vortex filaments per unit length is $\epsilon=\frac{\rho_s\kappa^2}{4\pi}\ln\left(\frac{R_\mathrm{cv}}{r_\mathrm{c}}\right)\sim 1.25\times10^{-12}\ \mathrm{J/ m}$. Thus, the energy injected per unit time is $2\pi {R_0} \times f \times \epsilon\sim 10^{-1} \ \mathrm{pW}$. The order of injected energy in this study is significantly lower than that in the experiments. However, it is numerically difficult to increase the power to make it comparable with the experiments.




\section{Conclusion}
We numerically investigate the emission of vortex rings from a statistically steady localized vortex tangle that was formed by colliding vortex rings. 
Our study begins the study of statistical laws of localized quantum turbulence.
We developed a system similar to the experiments and investigated the two statistical laws.
We succeeded in obtaining the laws, although the exponents in the power laws were different from the experimental results. 

In this study, although we performed simulations at $T=0$ $\mathrm{K}$, the experiments were performed at finite temperatures. 
An advantage of performing the simulation at $0\ \mathrm{K}$ is that, the vortex rings do not shrink spontaneously; hence, the sizes of the vortex rings emitted from the tangles can be easily fixed. However, at finite temperatures, the mutual friction shrinks the vortex rings whose sizes cannot easily determined. 

Therefore, it is ideal to perform the simulation at finite temperatures. There are predominantly two methods to achieve this. The first method is traditional, namely, following the vortex dynamics under the prescribed normal fluid velocity \cite{Schwarz1985}. This can be easily calculated, and we will consider in subsequent research. The second is to consider the fully coupling dynamics between a normal fluid component and a superfluid component \cite{Kivotides2000,Kivotides2007,Yui2018,Yui2020}. This is better than the first method. However, it is difficult to calculate the fully coupled dynamics if used for the present problem.

Our subsequent research will investigate the self-similar structure of vortex tangles, such as a fractal dimension \cite{Kivotides2001}, and associate with the statistical laws of the emitted vortex rings addressed in this paper. We will adjust the method used to inject vortex rings and investigate its effects on the statistical laws. For example, we can inject trains of vortex rings with expected turbulence generated by moving ions \cite{Walmsley2007,Walmsley2008}.
\begin{acknowledgments}
M. T. acknowledges the support from JSPS KAKENHI (Grant No. JP17K05548). H. Y. acknowledges the support from JSPS KAKENHI (Grant No. 15H03694).
\end{acknowledgments}

\bibliographystyle{apsrev4-2}
\bibliography{ref.bib}

\begin{thebibliography}{48}
\bibitem [{mem()}]{memo}%
  \BibitemOpen
  \href@noop {} {}\bibinfo {note} {The probability that a detecter catches more
  than one ring larger than $2R=10\ \mathrm{\mu m}$ during the period $[0\
  \mathrm{s}, 1.5\ \mathrm{s}]$ is about $2\%$ at most. Thus, a possibility
  that some detectors may catch more than one ring during the period is very
  low in our simulation.}\BibitemShut {Stop}%
\end{thebibliography}

\end{document}